# Chiral Plasmonics

Mario Hentschel[1], Martin Schäferling[1], Xiaoyang Duan[2,3], Harald Giessen[1], and Na Liu[2,3]

[1]*4th Physics Institute and Research Center SCoPE, University of Stuttgart, Pfaffenwaldring 57, 70569 Stuttgart, Germany*
[2]*Max Planck Institute for Intelligent Systems, Heisenbergstrasse 3, 70569 Stuttgart, Germany*
[3]*Kirchhoff Institute for Physics, University of Heidelberg, Im Neuenheimer Feld 227, 69120 Heidelberg, Germany*

**Abstract**

We present a comprehensive overview of chirality and its optical manifestation in plasmonic nanosystems and nanostructures. We discuss top-down fabricated structures that range from solid metallic nanostructures to groupings of metallic nanoparticles arranged in three dimensions. We present the large variety of bottom-up synthesised structures. Using DNA, peptides, or other scaffolds, complex nanoparticle arrangements of up to hundreds of individual nanoparticles have been realized. Beyond this static picture, we give also an overview over recent demonstrations of active chiral plasmonic systems where the chiral optical response can be controlled by an external stimulus. We discuss the prospect of utilizing the unique properties of complex chiral plasmonic systems for enantiomeric sensing schemes.

**Teaser**

We present a comprehensive overview of chirality and its optical manifestation in plasmonic nanosystems and nanostructures.

**Introduction**

Most generally, chirality refers to the symmetry properties of an object. Mathematically speaking, a chiral object does not possess mirror planes or inversion symmetry. The object and its mirror image can thus not be made to coincide by simple rotations or translations. One calls the object and its mirror image two enantiomorphs, or, in molecular systems, two enantiomers. As our own hands are chiral themselves, a chiral object is also termed "handed".

While this terminology sounds fairly technical, chiral objects are quite ubiquitous in nature. Apart from our own hands and feet, a large number of objects show chirality: The essential amino-acids, carbohydrates, nucleic acids, proteins, seashells, and the tendrils of climbing plants are examples from nature. In art and architecture chirality is an often seen feature in towers, columns, and staircases, to name a few. Also, a lot of daily objects are handed, such as corkscrews or keys. In fact, nature itself has been called chiral as many important and central aspects lead us to the so-called asymmetry of life: a large number of biomolecules only exist in one handedness, e.g., all 21 essential amino-acids are L-enantiomers, indicating that the associated physiological processes show 100 % stereoselectivity. Accordingly, chirality has been the subject of intense research investigations along the route to unravel the basic origin of life itself (*1*).

Apart from the purely geometrical considerations, chirality can also manifest itself in physical properties. In fact, symmetries are at the basis of the properties of nearly every physical system as they dictate the electronic structures, crystallographic ordering, chemical bonds, etc. (*2*). Chirality can for example manifest itself optically via a different response to right- or left-handed circularly polarized light (RCP and LCP), being the "chiral state" of light. A chiral medium exhibits different complex refractive indices for those two polarization states (*3*). The difference in the imaginary part results in a different absorption of RCP and LCP light, which is called circular dichroism (CD). A linearly polarized light field, on the other hand, will be rotated with respect to its original orientation while traveling through a chiral medium. This effect is called optical rotatory dispersion (ORD) and can be explained by the difference in the real part of the refractive index: A linearly polarized light field can be described as a superposition of LCP and RCP light. Due to the different refractive indices, a phase shift between the two orthogonal polarizations is induced. As a result, the plane of polarization will be rotated. Similar to the absorption and refractive index of an achiral medium, CD and ORD are Kramers-Kronig related (*4*).

CD and ORD spectroscopy are powerful tools in biology, medicine, chemistry, and physics for studying chiral molecules of different kinds and sizes, especially for analyzing the secondary structure and conformation of macromolecules (*5*). Structural, kinetic, and thermodynamic information of macromolecules can be readily derived from CD and ORD measurements. It should be noted that the chiral optical response of most biomolecules or other natural chiral media is generally very weak. High concentrations or large analyte volumes are required to study many of the aforementioned properties. Plasmonics, on the other hand, deals with the optical properties of metallic nanoparticles (*6*). The plasmonic resonances of metallic particles fundamentally stem from a large number of quasi-free conduction electrons. An impinging light field can induce a harmonic oscillation of these quasi-free electrons, which concentrate far-field radiation into subwavelength volumes. What is even more exciting is the extremely large dipole strength that is associated with these resonances. Particles of subwavelength sizes can have significantly large interaction cross-sections, which render individual particles extremely bright. Additionally, by virtue of the strong near-fields, resonances of individual particles can be coupled together and form collective modes that extend over the entire nanoparticle arrangement. In analogy to molecular physics, where the electronic wavefunctions of individual atoms mix and hybridize to form molecules, such plasmonic nanoparticle groupings have been termed *plasmonic molecules* in which plasmons from individual nanoparticles mix and hybridize. The related theory is called plasmon hybridization.(*7*) The last few years have seen a tremendous interest in such plasmonic molecules for creation of unusual optical properties and extended plasmon modes with modified properties as compared to the individual nanoparticles (*8*).

The field of chiral plasmonics can in fact be seen as an extension of these efforts. As described above, chirality is a very intriguing property in its very own right. In recognizing the weak chiral optical responses of natural systems on the one hand and the strong light-matter interaction in plasmonic systems on the other hand, the idea of enhanced and tailored chiral optical responses in chiral plasmonic systems holds great promises (*9*).

Most research in the field can be understood and classified following the same rules as in molecular systems. Nature offers a large variety of chiral molecules, with very distinct working principles. A hexahelicene molecule consists of a ring of six benzene units. Six such units constitute a full 360° turn. However, for such a hexahelicene molecule the ring is not closed in-plane, but cut and bent open. Depending on an upward or downward bending, the molecule can either be left- or right-handed. The molecule possesses delocalized electronic orbitals that extend over the entire structure. Another prominent example is the so-called chiral center in which the tetravalent bonds of a central carbon atom are dressed with four different atoms or molecular groups, rendering the structure chiral. One can imagine the structure as a symmetric three-sided pyramid whose four corners are each dressed with a different object. The unhanded symmetric pyramid therefore becomes chiral. Again, such a molecular structure is rendered chiral by the nature and symmetry of the interaction as well as the bond formation in the molecules. Therefore, they also exhibit a strong chiral optical response due to the handedness not only of the structure but also of the collective wave functions. As soon as the molecular entities become more complex and more individual subunits and compounds are involved, another very different working principle can be observed. For example achiral chromophores can be arranged in a handed fashion, e.g., helically attached to a central backbone. These chromophores can either have degenerate or nondegenerate energies (*3*). In the latter case the collective wave functions or orbitals of the subgroups only interact very weakly with the neighbouring ones, so that crosstalk is mainly mediated by Coulomb interaction between the charge clouds. A chiral optical response is in this case not caused by a collective handed orbital, but rather by an additional interaction between individual subgroups that are arranged in a handed fashion. The aforementioned chiral center is another example of this working principle: Many biomolecules do actually consist of several chiral centres which do not necessarily all share a collective orbital (*10*, *11*).

Chiral plasmonic structures can be classified in a very similar fashion, mainly depending on whether or not a collective plasmonic mode of handed character is formed along the entire structure. Collective plasmonic modes can form either in metallic nanoparticles which are themselves handed or in handed groupings of individual resonant nanoparticles. These two cases exhibit the strongest chiral optical responses and in fact represent the vast majority of chiral plasmonic structures in literature. Both concepts have been used in top-down fabricated systems, e.g., utilizing two-photon direct laser writing or electron-beam exposure, as well as in bottom-up synthesized structures, for example using DNA self-assembly. In analogy to biphenyl, one can also create handed arrangements of non-resonant (non-degenerate) plasmon modes. In such systems no collective modes of handed character are formed. However, a scattering type interaction is still possible which can lead to a chiral optical response (*12*). This response, as it is not associated with a handed plasmonic mode with large dipole strength, is much weaker.

One crucial difference between natural molecules and artificially created plasmonic molecules should be noted: a molecule is a bound system of individual atoms or molecular subgroups. Its structural symmetry and properties are determined by the interaction of the individual orbitals as well as by the symmetry and properties of the bond formation. Such a molecule therefore necessarily reflects its symmetry in the optical properties. A plasmonic molecule, however, is an artificially created entity that is not bound by the plasmonic interaction but by a physical scaffold that can be, e.g., a dielectric medium or DNA origami, among others. Thus, even though a plasmonic molecule might be handed from a geometrical point of view, there is no guarantee that the molecule will exhibit a chiral optical response. That said, a similar effect can be observed in molecular systems, in the extremely large size regime: As soon as a molecular system becomes so large that opposite ends are only weakly bound to each other they do not share a common electronic wavefunction. As a result the handed information of the structure is lost and the molecule shows extremely weak or no chiral optical response.

The following sections are structured as follows: first, we will discuss chiral plasmonic systems that are created using top-down fabrication methods, such as direct laser writing or lithography techniques. Subsequently, we will review the efforts of bottom-up synthesis using peptides or DNA-scaffolds. Different from natural molecules, it is fairly straightforward to manipulate the structure and arrangement of elements in artificial plasmonic molecules. This has led to the realization of active and tuneable plasmonic chirality, both in top-down and bottom-up structures, which we will discuss afterwards. We will end the review with a discussion on the potential applications of chiral plasmonics, particularly in view of the recent efforts in chiral sensing, that is, the ultimate goal of discriminating the handedness of single (or few) chiral molecules utilizing appropriate chiral plasmonic sensing platforms.

**Top-down fabrication methods for static chiral plasmonics**

An archetype chiral structure is a helix or a spiral, an object exhibiting a distinct twist and is well-known from daily life and from nature. Gansel and co-workers have demonstrated plasmonic helices, which were fabricated by two-photon direct laser writing and a subsequent electroplating step that filled the helical voids in the photoresist (*13*). The resulting free-standing helices exhibited plasmonic modes that extended over the entire structure, as can be seen in the simulated mode profiles shown in Figure 1A. The authors showed that the modes were of standing wave type, with increasing number of nodes, indicating higher order modes. Each of these modes had a distinct handedness and thus predominately interacted with light of the same handedness. The authors demonstrated that light of opposite, that is, non-matching handedness was blocked by the array of helices as the plasmonic antennas exhibited a very broad transmission minimum for LCP light while being highly transmittive for RCP light from 4000 nm to 8000 nm, cf. Figure 1A. The arrays therefore effectively behaved as a circular polarizer which converted linear incident light, which is a superposition of LCP and RCP light, into circular polarized light as the non-matching component was removed in the transmitted light field.

The same authors have studied the influence of pitch and diameter of the helices in great detail (*14*). A variation of the two parameters and the creation of tapered helices as shown in Figure 1B can further increase the operation bandwidth of the polarizer or in other words extend the spectral width of the respective plasmonic modes which underlines the direct connection of shape, plasmon modes, and chiral optical response. With impressive advances in fabrication technology the helices could be further optimized by nesting several helices into one another, as shown in Figure 1C (*15–17*). The authors used a lithography technique inspired by Stimulated Emission Depletion Microscopy to further increase the resolution of the two-photon direct laser process (*18*). It should be noted that this structural change has a major effect on the working principle of the polarizer: While for a single helix the suppressed transmission of one polarization is due to a difference in the reflection, the symmetry of three intertwined helices enforces similar reflectance for LCP and RCP. Thus, the observed transmittance difference is purely due to a difference in absorption, similar to chiral molecules where no scattering is observed. Another particularly intriguing structure has been described by Kaschke and co-workers (*19*). The authors stacked two helices of opposite handedness on top of each other. Due to the direct ohmic contact the helices are strongly coupled. The isolated upper spiral would block light of one handedness while transmitting the other polarization state. However, due to the strong conductive coupling between the two spirals the energy stored in the fundamental mode of the upper helix is efficiently transferred into the fundamental mode of the lower helix, which is then radiated to the far-field. As the modes have opposite handedness the combined structure thus acts as a circular polarization converter.

Apart from direct laser writing other techniques have been used to create chiral plasmonic structures. Frank and co-workers used colloidal nanohole lithography, a technique at the boundary of top-down and bottom-up techniques, to create gold spirals or ramps (*20*). Figure 1D depicts an exemplary tilted overview and close-up image of the structures. The authors observed a number of plasmonic modes which extended over the entire nanostructure, thus leading to the strong chiral optical response. Using glancing angle deposition a number of researchers have demonstrated high quality plasmonic spirals (*21*). Mark and co-workers showed two-pitched gold nanohelices with a strong chiral optical response around 700 nm (*22*). The TEM images and the mirrored CD spectra for the two enantiomers, as shown in Figure 1E, underline the excellent quality of the structures. Using a similar technique, Bai and collaborators observed a strong CD response in silver nanohelices at even shorter wavelength in the blue spectral region (*23*). The overall scheme can also be reversed by drilling a "solid" handed aperture into a metallic film leading to an "inverse" spiral structure (*24*). Höflich and co-workers demonstrated that electron-beam induced deposition could be used to produce plasmonic spirals (*25*). While the technique is highly versatile and allows to directly write functional structures, most structures suffer from rather low material quality. The precursor gases used in the deposition often cause carbon and other contaminations which harm the material quality. However, Esposito et al. have recently demonstrated plasmonic spirals that exhibited multiple plasmonic modes and pronounced chiral optical signatures in the visible regime, indicating excellent material quality (*26–28*). An array of such structures is depicted in Figure 1F.

The aforementioned structures are real spirals, some of them even exhibiting several pitches. Another possibility is to create structures that are not directly reminiscent of a spiral but do still possess a handed character, such as staircases or a Möbius strip (*29*). Dietrich and co-workers used the so-called on-edge lithography to create a three-dimensional L-shaped metallic strip, as can be seen from the close-up scanning electron micrograph shown in Figure 1G (*30*). Strong plasmonic modes extending over the whole length of the structure could be observed (*31*). Using a similar idea, Dietrich and co-workers evaporated gold through a star-shaped aperture onto cone-shaped posts. The obtained 3D chiral structures also exhibit a strong CD response (*32*). Another related ansatz by Yeom et al. utilized tilted evaporation of gold onto hexagonally shaped dielectric posts leading to a handed metallic film on this platform exhibiting a CD response in the visible wavelength region (*33*, *34*). Helgert et al. have used very sophisticated multi-step electron beam lithography to create a handed metallic nanostructure fused from two individual L-shaped nanoparticles exhibiting a CD response in the near infrared spectral region (*35*). Anther interesting concept has been introduced by McPeak and co-workers. They utilized different etching rates in silicon to fabricate complex indentations in high index wafers. By pre-structuring a resist film on the wafer they demonstrate remarkable shape control of the indentations. Using a sophisticated evaporation and lift-off technique they could transform the indentations into solid gold nanoparticles with distinct structural handedness. The resulting nanoparticle solutions exhibited a strong chiral optical response in the visible and near-infrared spectral range.

It needs to be stressed that all the mentioned structures can in fact be viewed as intrinsically chiral structures as they are all made of metals and exhibit a distinct handedness. Therefore they exhibit strong chiral optical responses. The formation of chiral plasmonic modes is intrinsically ensured. It should also be noted that many deformed particles and shapes can in fact be chiral: Spheres, rods, or similar with specifically designed (or random!) indentations or deformations can lead to a chiral geometry which is also intrinsically handed (*36*). Many of these chiral particles do not at all resemble spirals or similar systems, but can still not be superimposed with their mirror images – they are sometimes colloquially called "potato-shaped". The major difference between these two classes is that it is usually straightforward to tell the handedness of objects from the first class (such as a helix), while this is not the case for potato-shaped structures (*37*). Whether or not such deformed particles exhibit a strong chiral optical response is somewhat complicated and is also largely determined by the extent of the deformation, which in turn influences the deformation of the plasmon mode excited in the nanoparticles. Many plasmon modes, despite irregularities in their shape, still very well resemble a simple dipole.

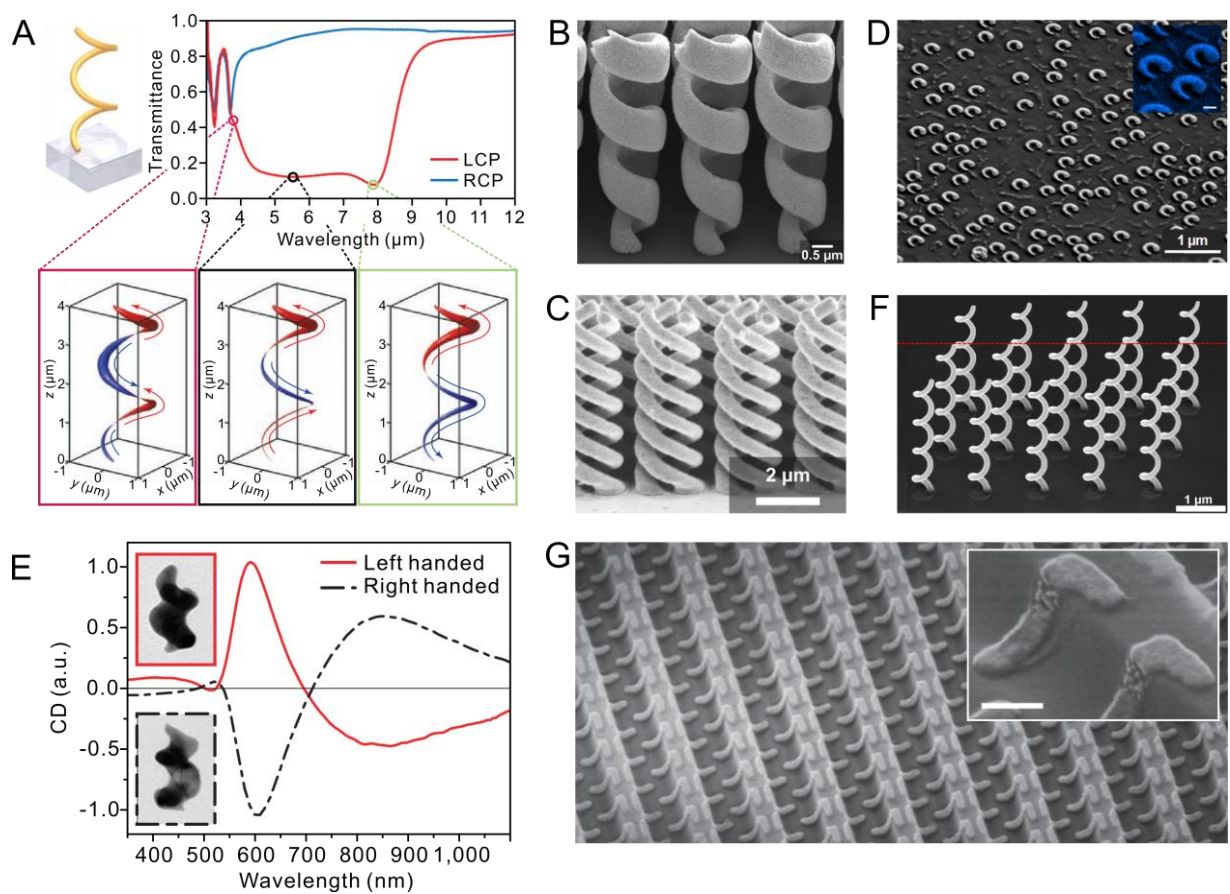

**Figure 1**: Solid plasmonic chiral structures. All these structures are intrinsically chiral, meaning they are all made of solid metal and exhibit a distinct handedness due to their shape. A) Simulated transmittance spectra and plasmonic modes of a metallic spiral with two pitches. The structures basically blocks LCP light while transmitting RCP light nearly without loss. The observed plasmonic modes extend over the entire structure thus being strongly handed themselves. B) Tapered gold helices. In manipulating pitch and diameter of the spirals, the plasmonic modes can be manipulated and an even larger operational wavelength band can be generated. C) Nested plasmonic helices fabricated via STED direct laser lithography. D) SEM image of plasmonic spirals fabricated via colloidal nanohole-lithography. Being at the boundary of top-down and bottom-up techniques, this lithography technique allows for large area fabrication. E) GLAD (glancing angle deposition) allows for the fabrication of extremely small solid gold spirals exhibiting a strong CD response in the visible wavelength range, with excellent mirror symmetry of the CD spectra of the two enantiomers, as expected from theory. F) Free-standing metallic spirals fabricated by electron beam induced deposition, which allows to directly write functional nanostructures. G) A two-step lithography process allows fabricating on-edge solid metallic L-shapes which show a strong chiral optical response. Figures reproduced with permission from… A,(*13*) B,(*14*) C,(*18*) D(*20*), E,(*22*) F,(*26*) G.(*30*)

The majority of micro- and nanostructuring is lithography based, i.e., manufacturing was carried out by optical or electron beam lithography, which is generally associated with a planar and layered process. As researchers in plasmonics have mostly being using these lithographic techniques, the majority of chiral plasmonic structures have been fabricated in layers, meaning they consist of several functional layers that are stacked on top of each other in order to render the structure three-dimensional, a prerequisite for a structure to be chiral. Consequently, all these systems are plasmonically coupled systems. In order to show modes that extend over the entire structure, resonant coupling is crucial.

Conceptually, it is possible to arrange achiral constituents to form a superstructure that is chiral. The most straightforward design consists of two individual nanorods that are stacked on top of each other and are twisted with respect to each other, see Figure 2A (*38*). Despite its simplicity, this system has a number of degrees of freedom, including the twisting angle, the lateral alignment, and the vertical spacing distance. In general, the optical response is complicated due to the presence of two modes whose resonance frequencies and amplitudes sensitively depend on the relative orientation as well as the excitation direction (*39*, *40*). Yin et al. have shown that two orthogonally coupled nanorods can be viewed as the plasmonic version of the so-called Born-Kuhn model (*41*). The coupled systems exhibits two modes, the symmetric and the antisymmetric one, each exhibiting a distinct handedness. If the vertical coupling distance is matched to the resonance wavelength of the system, each mode is excited solely by RCP or LCP light. Figure 2A illustrates this mode behaviour. As a result, the observed CD spectrum shows one single dispersive lineshape, cf. Figure 2B, which is well known from chiral molecules. Knowing the CD response of the system, it is possible to calculate the optical rotatory dispersion (ORD) as the two quantities are Kramers Kronig related. This finding also indicates that in many studied systems a multitude of individual modes can be excited and lead to complicated CD spectra that are not always straightforward to interpret. Zhao and co-workers have studied the response upon stacking multiple rod layers on top of each other, rendering the mode formation more intriguing (*42*).

Other than nanorods, basically every plasmonic element can be utilized as a fundamental building block. Over the years a variety of structures have been demonstrated, using gammadions of different shapes (*43*, *44*), split ring resonators (*45*, *46*), direct and inverse crosses (*47*, *48*), L-shapes (*49*, *50*), and others, see Figure 2. In particular, the more complex building blocks, such as gammadions, support several plasmonic modes already in the individual building blocks, rendering the resulting structures somewhat broadband in their operation. Additionally, due to their physical size and volume, the resonance dipole strength tends to become very large allowing for very strong chiral optical responses, for example high gyrotropy as shown by Rogacheva and co-workers (*43*). Such structures have also been discussed in the framework of negative index matermaterials, see Figure 2C (*44*). Split ring resonators as well as crosses have been used to unravel the plasmonic coupling between the individual building blocks as well as the nature of the plasmonic modes (*46*, *47*). Decker and co-workers also pointed out the importance of the grating and periodicity in the systems (*45*). While molecules in a solution do not show any relative order and therefore lead to isotropic media, top-down fabricated structures are always restricted on an interface. The authors stressed that the arrangement of the individual nano-objects on the surface can lead to a biaxial system in which the refractive indices for light polarized along these axis are different. Therefore, the systems exhibit strong polarization conversion phenomena, which can mask, influence, or appear as a chiral optical response. In terms of symmetry and modes, the arrangement of the individual structures needs to be such that the eigenpolarizations are circular. Examples include the in plane isotropic symmetry groups $C_3$ or $C_4$, as is nicely illustrated in the scanning electron micrographs shown in Figure 2D. This effect, arising from the nature of top-down techniques, is always present in top-down fabricated systems and can be seen as a unique feature, different from natural chiral molecules.

If one uses nanospheres or nanodisks as simplest building blocks, one has generally two possibilities to render an arrangement chiral, as illustrated in Figure 2F (*51–53*). One can arrange identical objects in a handed fashion, termed configurational chirality. Alternatively, one can arrange different objects in an unhanded fashion, for example at the tips of a symmetric pyramid (which is in fact the analogy of the aforementioned chiral center) or at the corners of a square (see Figure 2G) (*53*) termed constitutional chirality. In doing so, chirality and any resulting chiral optical response cannot be explained by the symmetry of the individual building blocks, but rely on their three-dimensional arrangement. It is important to note that in both cases plasmonic coupling or cross talk is of crucial importance. The formation of collective plasmon modes that extend over the entire structure, similar to the case of more complex building blocks, is crucial. In fact, it has been shown that a strong chiral optical response cannot be obtained once the individual building blocks are strongly detuned (*52*). It should be noted, though, that plasmons can also interact via scattered radiation and via Coulomb interaction and not only via near-field coupling (*51*). This phenomenon is very similar to the case of biphenyl where the chiral response is not induced due to the formation of shared orbitals but due to Coulomb interaction. It has recently been shown that such a phenomenon can also be obtained in off-resonantly coupled plasmonic nanostructures in which indeed no collective modes are formed but off-resonant scattering in between the building blocks leads to a chiral optical response (*12*). However, as expected, the response is significantly weaker as compared to that of the resonant systems.

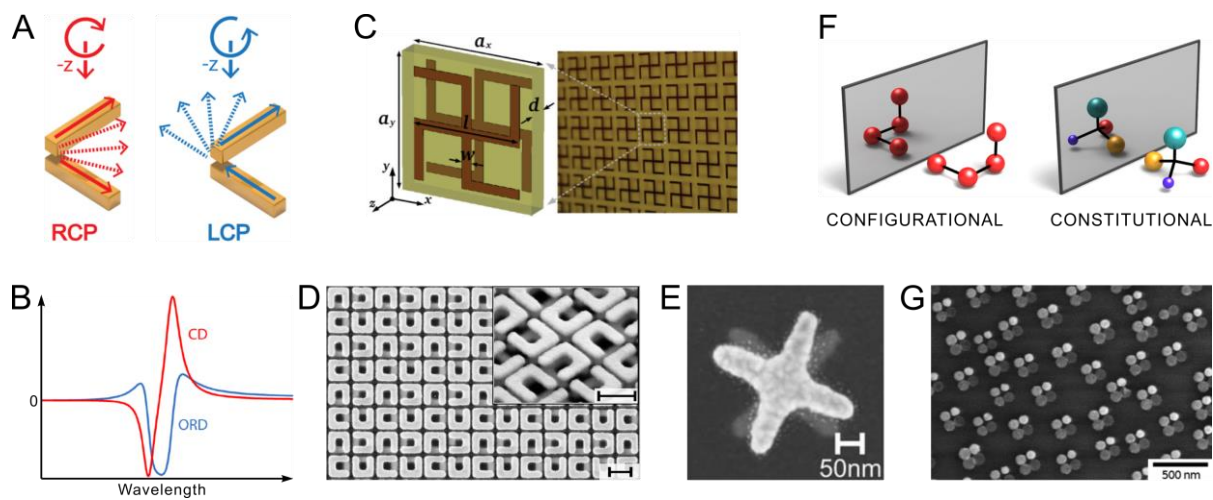

**Figure 2**: Chiral assemblies of achiral and chiral building blocks. A) Two bars stacked on-top of each other can be viewed as a fundamental chiral building block and as an analogy to the well-known Born-Kuhn-Model. The assembly supports two modes which, if ideally tuned, lead to a single dispersive lineshape in the CD spectrum, shown in B). The ORD spectrum can be calculated from the CD spectrum as they are Kramers Kronig related. C) Two-layered chiral structures composed of two individual 2D chiral gammadion shapes. D) Stacked and twisted split ring resonators arranged in a $C_4$ symmetric lattice in order to render the eigenmodes of the system truly chiral. E) Twisted stacked crosses. F) Assemblies of achiral nanoparticles can be rendered handed due to configuration or constitution. G) Illustration of constitutional chirality: Four particles of different height (as can be seen due to their different brightness) in an achiral arrangement generate a handed system. Figures reproduced with permission from… A,B*(41)*, C,*(44)* D,*(45)* E,*(47)* F,*(52)* G.*(53)*

**Bottom-up fabrication methods for static chiral plasmonics**

For large scale and solution-based synthesis of chiral plasmonic structures self-assembly has been a particularly powerful route. In general two different strategies have been used: Metallic nanoparticles can be organized with the help of molecular scaffolds that take the form of fibers and possess either an intrinsic twist or are applied in twisted layers. The other option uses self-assembly via self-terminated interaction that lead to chiral clusters, that is, individual plasmonic molecules. Due to the tailorability of the molecular interaction this route tends to show a larger degree of structural control over the end products.

Guerrero-Martinez and co-workers used anthraquinone based oxalamide fibers infiltrated with gold nanorods (*54*). In solution, the fibers took a twisted form so that the gold particles, as they were attached to the fibers, exhibited a helical arrangement. (see the TEM image in Figure 3A). A scattering type or Coulomb interaction between the individual particles led to the observed chiral optical response (*51*). In a related ansatz Song et al. used peptites to organize gold nanoparticles on a double helical scaffold (*55*). The resulting chiral nanoparticle chains reached considerable length with densely spaced nanoparticles facilitating coupling between the individual particles. The overall system showed a strong chiral optical response, similarly to the previous study, in the visible wavelength regime, see Figure 3B. Jung et al. used so-called modular gelator components which were self-assembled into stacks and thus formed a handed hydrogel scaffold for the subsequent growth of gold nanoparticles (*56*). Due to the even smaller sizes, the chiroptical response shifted to shorter wavelengths. Another very interesting motif was introduced by from Querejeta-Fernandez and co-workers (*57*, *58*). The authors used self-assembled cellulose nanocrystals loaded with gold nanoparticles, as schematically illustrated in Figure 3C. The nanocrystal itself took the form of stacked layers of fibers twisted with respect to one another. The particles immersed into the crystal were oriented along the fiber axis in the twisted layers and thus also formed a handed system. All the aforementioned examples are large scale handed plasmonic systems. However, despite their beauty and ingenuity, they offer little to no distance control in between the particles and thus only allow tuning of the plasmon coupling strength via particle concentration and thus the average inter-particle distance which, in particular in the fiber based strategies, tends to be rather large.

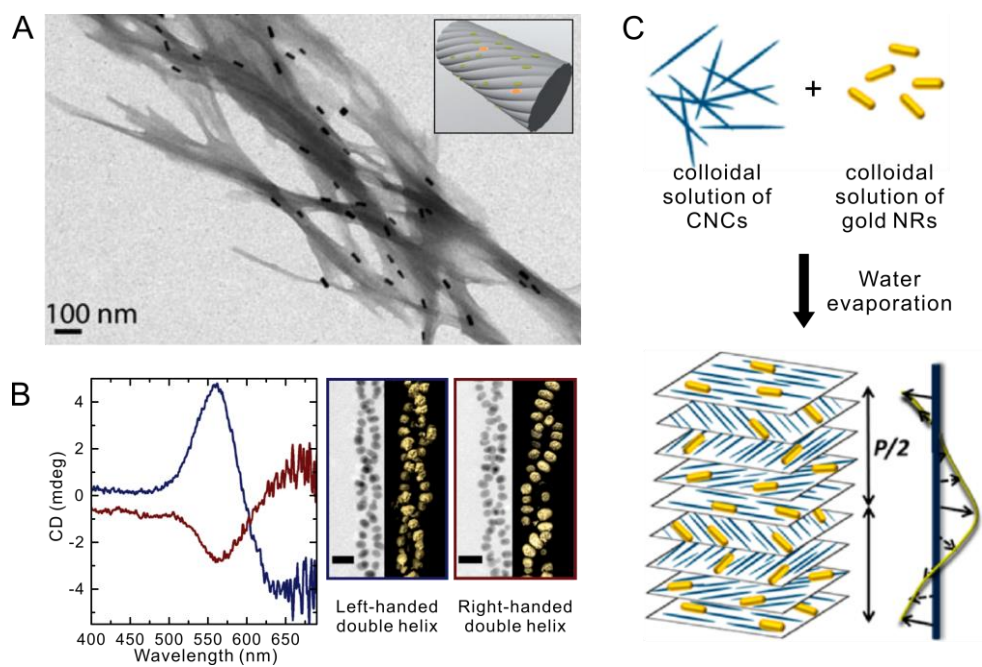

**Figure 3**: Chiral plasmonic structures assembled on scaffolds. A) TEM image of twisted anthraquinonebased oxalamide fibers which are infiltered with gold nanorods. B) CD spectra of double helix assembled using peptides. The structure is illustrated with the help of 3D surface renderings of topographic volumes. C) A system of layered twisted cellulose nanocrystals is used as a scaffold for gold nanorods. Figures reproduced with permission from... A,(*54*) B,(*55*) C.(*58*)

The very powerful bottom-up strategy via DNA self-assembly has been introduced many years ago. In general there are two working principles, DNA-origami and DNA-scaffolding. The first one uses templates, such as sheets, constructed from long DNA strands. The origami is dressed by designable binding sites, composed of single stranded DNA (ssDNA) for attachment of nanoparticles which are functionalized with complementary ssDNA strands through DNA hybridization. The latter method uses single ssDNA strands on individual nanoparticles that facilitate the binding between two specifically chosen nanoparticles or a group of particles such that they assemble only following the designed scheme. Both methods have a very high degree of tunability, programmability, and control over particle number, position, and sequence. Kuzyk and co-workers demonstrated an gold nanoparticle helix structure constructed from a sophisticated origami template (*59*). A DNA bundle made from 24-helices was helically dressed with nine specific binding sites. The addition of gold nanospheres functionalized with complementary DNA strands led to the formation of plasmonic helices with an impressively high yield. The assemblies showed a chiral optical response around 550 nm, see Figure 4A. The authors demonstrated an enhancement of the chiral optical response by overgrowing the gold nanoparticles with silver shells. The CD spectra shown in Figure 4B demonstrate that due to the stronger coupling between the individual nanoparticles the strength of the chiral eigenmodes of the helices increases. This bottom-up result further underlines the importance of efficient coupling and hybrid mode formation in these plasmonic nanostructures.

Shen and co-workers assembled a similar helical structure, yet with an intriguing strategy using an origami sheet (*60*). After dressing the origami sheet with two linear chains of the helper strands (ssDNA), gold nanoparticles functionalized with complementary strands were subsequently attached. The origami sheet was rolled up into a cylinder by the folding strands, which resulted in a two-pitched gold nanoparticle helix, exhibiting a chiral optical response around 550 nm, close to the resonance of the individual nanoparticles. TEM images of the resulting structures are shown in Figure 4C.

Shen and co-workers also assembled chiral structures composed of two nanorods aligned tip-to-tip under an 90° angle, schematically illustrated in Figure 4D (*61*). The attachment sites for the two nanorods were attached on opposite sides of the nanosheet. The position, orientation, and relative angle between the nanorods can be precisely controlled by the attachment sites on the nanosheet (*62*). Lan et al. used DNA nanosheets to create stacks of rotated gold nanorods (*63*). The individual sheets were dressed with capture strands on both sides, forming an 'X'-shape, as shown in Figure 4E. The strands were both complementary to the ssDNA used to assemble the gold nanorods. Mixing both entities led to the formation of nanorod stacks, exhibiting a chiral optical response around 700 nm. A distribution of finite stacks forms, as indicated by the TEM images in Figure 4E.

Drawing inspiration from molecular physics, the analogy to a chiral center, as introduced above, has been studied extensively as well. Here, four different particles are arranged in a symmetric frame in order to form a chiral entity. First results were demonstrated by Mastroianni and co-workers (*64*) as well as Chen and co-workers (*65*) using DNA scaffolds, that is, single and complementary ssDNA on individual nanoparticles lead to rationally designable molecules. Using the same concept, Yan et al. have created nanoparticle pyramids constructed from two differently sized gold nanoparticles, a silver nanoparticle as well as a quantum dot (*66*). Due to the specific interaction between the DNA strands the authors of both studies demonstrated precise control over the position and sequence of the nanoparticles, as illustrated by the schematic scheme shown in Figure 4F. However, only the latter study could demonstrate a chiral optical response in the blue spectral region. One possible explanation lies in the rather inefficient interaction between the individual nanoparticles as the resonance energies and the resonance dipole strength are highly detuned. In any case, the overall structure is highly tunable and the chiral optical response can even be inverted by appropriate symmetry breaking (*67*). Shen and co-workers could indeed show how to increase the chiral optical response in such a four-particle self-assembled geometry (*68*). They used an origami sheet, here with four docking sites, three along an L-shape on the one interface, one on the other, as shown in Figure 4G. Addition of identical-sized gold nanoparticles led to the formation of a chiral quadrumer. The handedness depended on the position of the fourth particle relative to the L-shape and could be freely chosen in the design process of the origami. In this structure, the handedness was induced due to the geometrical placement of the particles (configuration) rather than by size (conformation). In the experiment, depicted in Figure 4G, one observes a strong chiral optical response due to the efficient plasmonic coupling of the resonant particles and the formation of a chiral mode extending over the entire handed structure, such also being handed. Urban and co-workers used four curved DNA origami monomers to create an origami ring that was dressed with a total of 24 gold nanoparticles which were arranged in a spiral fashion around the ring (*69*). The resulting structure constitutes a plasmonic toroidal structure with a strong chiral optical response.

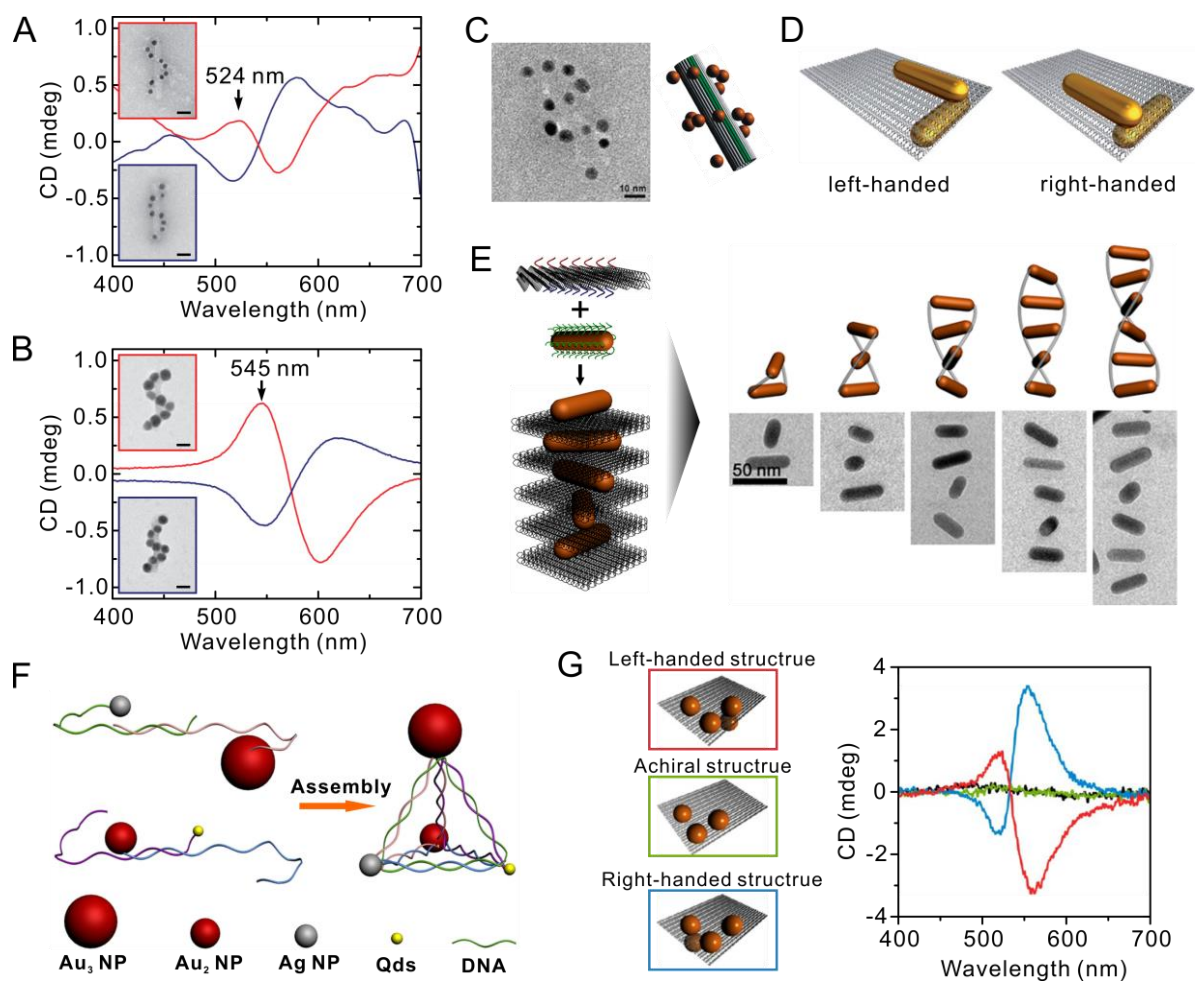

**Figure 4**: Chiral plasmonic structures assembled with the help of DNA strands and DNA origami. A) A DNA backbone with helically attached binding sites is used to create helices of gold nanoparticles exhibiting a CD response around 550nm. B) Overgrowth of the particles with silver leads to an increased coupling strength, a much stronger CD response, and a small spectral red shift due to the increased coupling. C) Gold nanoparticles are arranged in two parallel rows on a DNA origami and subsequently rolled up to form a spiral. D) Utilizing the possibility to differently facilitate the two sites of a DNA origami sheet, a handed rod-dimer can be assembled. E) Origami sheets are used to create stacks of rotated gold nanorods. The individual sheets are dressed with capture strands on both sides and adding gold nanorods which are dressed with complementary ssDNA to the formation of nanorod stacks. F) Using specifically designed ssDNA nanoparticle pyramids constructed from two differently sized gold nanoparticles, a silver nanoparticle, as well as a quantum dot can be assembled. G) An origami sheet with four docking sites, three along an L-shape on the one interface, one on the other is used to assemble a chiral quadrumer. The handedness depends on the position of the fourth particle relative to the L-shape. The structures show a strong CD response. Figures reproduced with permission from… A,*(59)* B,*(59)* C,*(60)* D,*(61)* E,*(63)* F,*(66)*  G.*(68)*

**Active chiral plasmonics**

The handedness of a molecule is a fixed and well defined property. Fundamentally speaking, this is the reason why CD spectroscopy is an important tool for structural investigations as the spectral response is uniquely related to a certain handedness or to a single enantiomer of a system. In chiral plasmonics, researchers have demonstrated excellent structural control, both using top-down and bottom-up techniques. With such control, systems with switchable chirality can be realized, i.e., systems that show different chiral optical responses under a certain external stimulus. In molecular physics a change in the chiral optical response is always related to a real change in the 3-dimensional arrangement, hence, the geometry of a system. In artificial structures it is not straightforwardly possible to change the arrangement of the structures post-fabrication.

However, changing the chiral optical response is still possible by any manipulation that suitably influences the plasmon modes of the system, even if the geometrical handedness of the structure stays unchanged. Zhang and co-workers demonstrated a highly sophisticated top-down structure which showed a photoinduced optical handedness change (*70*), see Figure 5A. Using multi-layer lithography they created a three-dimensional molecule composed of two individual chiral units that together defined the optical response of the entire system. The incorporation of silicon pads allowed to optically address the chiral units and to manipulate their relative contribution to the overall response. Under illumination with an external light field the optical response changes without a structural reconfiguration of the chiral molecule. Yin and co-workers showed a similar system, shown in Figure 5B, using a phase change material (GST-326) as the switchable medium, allowing for a spectral shift as well as a sign change of the optical response (*71*). GST exhibits a large refractive index change when the material is thermally switched from its amorphous state into the crystalline state. The authors demonstrated a very large spectral shift of the chiral optical response of a two-bar system utilizing this active thermal switching mechanism. Additionally, they demonstrated the switching of the chiral optical spectrum by using a similar bias-structure concept as in the previously mentioned work. Two chiral units, one active chiral dimer and one bias dimer, generated the overall chiral optical response. One of them could be actively detuned with help of the GST layer which lead to a pronounced sign change in the optical response.

A similar tuning concept was introduced by Duan and co-workers (*72*). They used a hybrid chiral plasmonic system that consisted of individual gold and magnesium nanoparticles in a gammadion-like arrangement, shown in Figure 5C. Hydrogen allowed for switching the magnesium to magnesium hydride, which is a dielectric. Consequently, the plasmon modes of the magnesium particles vanished. With these modes no longer present, the near-field coupling in the entire systems broke down. The remaining plasmon modes in the gold nanoparticles were no longer chiral. The chiral optical response was turned off. The structure thus enabled a continuously tunable chiral optical response depending on the hydrogen concentration within the magnesium.

In all the aforementioned systems, the change in the chiral optical response is enabled by sophisticated designs and is a consequence of the superposition of different signals. Schreiber et al. demonstrated a switchable system that relies on the fact that the excitation of plasmon modes strongly depends on the orientation of the nanostructure (*73*). Plasmonic helices, assembled using DNA origami templates (*59*), were tethered to a surface in solution. In solution the helices stood upright. After removal of the buffer solution and drying of the sample the helices lied flat on the surface. This change of the relative orientation between the helices and the incident light lead to a switching of the chiral optical response.

Beyond these methods to manipulate the plasmon modes, the post-fabrication tunability of bottom-up structures additionally allows for actively and reversibly changing the geometrical arrangement and configuration of a chiral plasmonic molecule, that is, its handedness, and consequently fully inverting the chiral optical response. The first fully switchable system has been shown by Kuzyk and co-workers: The authors constructed an origami template which consisted of two intersecting twisted bundles held together, shown in Figure 5D. The tunable angle between the two bundles was controlled by two DNA locks (*74*). These locks could be locked or unlocked with the help of specifically designed fuel DNA strands. Two gold nanorods were attached on the two bundles. Addition of the fuel DNA strands as the external stimulus allowed to controllably and reproducibly switch the twisting angle between the rods and thus change the handedness of the structure, leading to a flipped chiral optical spectrum. The stimulus that triggered the structural and thus the optical change was differently sequenced DNA strands that mediated the opening or closing of the DNA locks. While highly interesting, this is obviously a challenging experimental technique. The same group of authors presented a different design that uses ultraviolet and visible light illumination as the input trigger. A similar structure of two origami bundles with two gold nanorods was locked in a twisted and therefore handed geometry by an azobenzene-modified lock. Under UV (VIS) illumination, the molecule underwent a trans-cis photoisomerization locking or unlocking state of the two DNA bundles, thus rendering the gold nanoparticle pair either handed or achiral. The strong chiral optical signal in the locked state vanished under UV illumination due to formation of the relaxed state and reappeared under VIS illumination due to formation of the locked state.

Another concept uses DNA origami with individually addressable binding sites (*75*). Two nanorods were attached under a 90° angle on the opposite sites of a DNA origami sheet, shown in Figure 5 E. One was static on the sheet, the other one could walk over the surface of the sheet with the help of fuel strands interacting with 6 specific binding sites arranged along the track. Upon addition of the fuel strands, the DNA strands covered on the walker nanorod were released from the previous binding sites and were attached to the next binding sites successively, thus walking along the track. The arrangement of the two nanorods could in this way be manipulated in five distinct steps switching the handedness of the system from left to right and back. Along with the structural change, the spectra showed distinct changes and an ultimate full reversal of the spectrum. In creating an even more complex walker landscape out of DNA the authors demonstrated the ability to deduce structural changes and arrangement on the nanoscale from the optical response alone. The same group of authors extended the concept by incorporating a total of three rods . They showed that the optical response could not only be fully controlled with the help of an external stimulus but also that is also possible to optically resolve the movements of two separate and individually addressable walkers on the DNA origami (*76*).

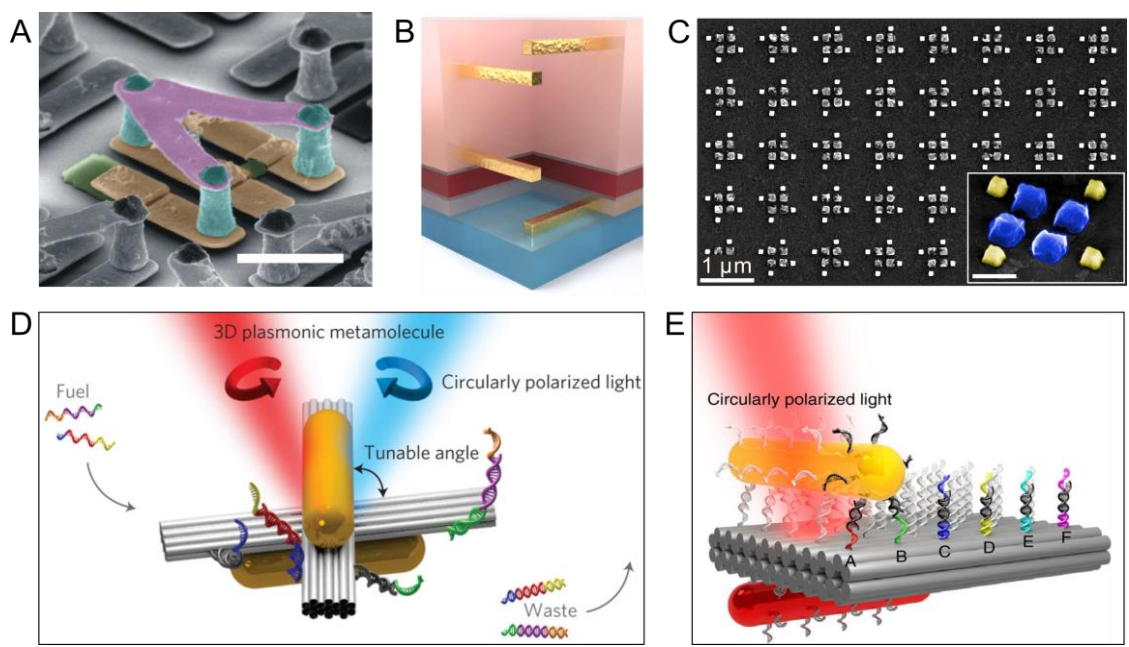

**Figure 5:** Switchable chiral plasmonic structures. A) Top-down fabricated three-dimensional molecule consisting of two individual chiral units which together define the optical response of the entire system. The incorporation of silicon pads (green) allows to optically address the chiral units and to manipulate their relative contribution to the overall response. B) Two chiral units, one active chiral dimer (bottom) and one bias dimer (top), generate the overall chiral optical response while one of them can be actively detuned with help of the GST layer, leading to a pronounced sign change in the optical response. C) Switchable hybrid chiral plasmonic system, consisting of gold and magnesium nanoparticles in a ratchet wheel-like arrangement. Switching the magnesium plasmon off via hydration switches the chiral plasmonic response off. D) Interlocked DNA bundles are dressed with two gold nanorods. The locks can be unlocked with the help of specifically designed DNA strands, followed by the formation of new DNA bonds that lock the two bundles with the opposite twisting angle thus inverting the handedness and the chiral optical response. E) Schematic illustration of DNA nanowalker. The lower red rod is static while the upper yellow one can walk over the origami sheet successively switch the chirality of the system. Figures reproduced with permission from… A,(*70*) B,(*71*) C,(*72*) D,(*74*) E.(*75*)

**Chiral Sensing**

As shown, one of the main striking features of plasmonic chirality is its strength compared to naturally occurring chiral molecules. Even dense solutions of chiral molecules might show a CD response of only a few milidegrees. ~~Accordingly, the so-called dissymmetry factor is only in the range of $10^{-7}$ to $10^{-5}$.~~ Plasmonic molecules and structures, on the other hand, can exhibit CD responses of tens of degrees for structures less than a wavelength in thickness. In the extreme case, this can result in responses where one handedness is fully absorbed while the other is transmitted (*77*). ~~The resulting dissymmetry factors approach unity.~~ The reason, as discussed above, is not surprising. In contrast to molecules, plasmonic structures are excellent scatters and absorbers for light.

It is an intriguing question whether it is possible to combine plasmonic and molecular systems such that the chiral optical response of the chiral molecule will be increased. Such hybrid chiral systems could result in novel chiral optical spectroscopy techniques with largely increased sensitivity, similar to plasmonically enhanced spectroscopy techniques for achiral molecules such as SERS or SEIRA (*78*, *79*). The field of plasmonically enhanced chiral optical spectroscopy has been very active because of the high potential benefit for both fundamental physical and chemical research as well life science applications. However, it has also been a field of intense discussions as many of the aspects of the assumed interaction and processes are not yet well understood.

Many studies rely on the interaction of an achiral plasmonic resonance with chiral molecules. The plasmonic resonance can either result from one single achiral nanostructure or in an achiral assembly of several nanostructures. In the following, we will discuss some important findings. More comprehensive reviews can be found in Ref. (*80*, *81*).

In such a hybrid chiral system, two distinct features can be observed: Firstly, the intrinsic CD of the molecule itself is modified. Secondly, an additional feature in the CD spectrum appears at the plasmon resonance wavelength. This second effect shifts the CD response from the UV, where it generally occurs in molecular systems, more into the visible wavelength region around the plasmon resonance. Using plasmonic nanostructures with resonances in the UV or blue wavelength region increases the strength of the effect (*82*, *83*). Govorov and co-workers interpret the observed phenomena as a Coulomb interaction (*84*). Several experiments reported both the modified CD response as well as the induced signal at the plasmon resonance, using gold nanoparticles and peptides (*85*), silver nanocrystals and cysteine (*86*), or gold nanoparticles and tobacco mosaic viruses (*87*), underlining the generality of the effect. A first definitive proof of the importance of plasmonic coupling was reported by Maoz and co-workers (*88*). They introduced molecular spacer layers of well-controllable thickness between the plasmonically active particles and the chiral analyte which lead to a suggestive decrease of the induced CD signal, as expected for a near-field interaction. Abdulrahman and co-workers have further shown that induced chirality can also occur in a far-field coupling scheme, similar to the coupling of purely plasmonic systems (*89*). It should be noted that the enhancement of the chiral optical response and the strength of the induced CD are rather weak and will not exceed roughly one order of magnitude if only one close-by nanoparticle is considered. Stronger enhancements can be observed in the hot-spots of nanoparticle assemblies. Gerard and co-workers studied an aggregated system of gold nanoparticles and oligonucleotide molecules (*90*). Only in case of the formation of small gaps an induced CD signal could be detected. Govorov and Zhang could further confirm that the effect depends strongly on the field enhancement in the gap and is also highly sensitive to the relative orientation of the molecular dipole and the dimer axis (*91*, *92*).

More complex geometries of the nanostructure result in even more sophisticated chiral interactions, which cannot be fully described by the aforementioned dipolar coupling schemes. Tang and Cohen introduced a measure for the chiral interaction strength of an arbitrary electromagnetic field, which they termed optical chirality (*93*). Complex plasmonic nanostructures were optimized to generate near-fields with high optical chirality (*94*), where the best results could be obtained for chiral plasmonic nanostructures (*95*). However, geometrical chirality is not mandatory; even achiral plasmonic nanostructures can generate chiral near-fields locally (*96*, *97*). This behaviour constitutes a major difference between the chiral optical far-field and near-field response. An important conclusion is that the CD of a nanostructure cannot be used as a measure when designing a structure with strong a structure with strong chiral near-fields and vice versa (98).

Although the concept of optical chirality offers a powerful tool to quickly check different designs regarding their potential usefulness for plasmonically enhanced chiral sensing, the theory assumes no back action of the chiral molecule on the nanostructure and, therefore, cannot fully predict the response of such hybrid systems. Rigorous numerical simulations that include the chiral medium close the gap because they take all electromagnetic interactions into account. The results of such simulations indicate that strong field enhancement and the correct alignment of the enhanced fields with respect to the incident field is more crucial than the chirality of the nanostructure (*99*). Some experimental investigations have been performed by Hendry and co-workers, who used a gammadion-shaped nanoparticle to study the interaction of such complex nanostructures with chiral molecules (*100*). They claimed a sizable resonance shift of the chiral optical response of the metallic nanostructure when adding the chiral analyte. Molecules with different handedness resulted in different signs of the shift. The strength of the response could not completely be explained by theoretical considerations, indicating that the interaction mechanisms in such complex hybrid systems have not been fully revealed yet.

However, a different approach for ultra-sensitive detection of chiral molecules has been discussed in literature during the last years. It utilizes the strong CD response of chiral plasmonic systems, which has been extensively discussed in this review, in a highly ingenious way. In this approach, the addition of the analyte triggers a self-assembly process which results in coupled plasmonic structures that exhibit a CD response. The emergence of the plasmonic CD response, which can be traced with standard CD spectrometers, therefore reports the presence of the analyte. Several systems have been studied. For example, Wu and co-workers used the formation of nano-particle dimers for the detection of microcysteine-LR and a cancer biomarker (*101*). Similar mechanisms have been used for the detection of silver ions(*102*) and bisphenol A (*103*). Zhao et al. used shell-engineered systems for DNA detection (*104*). In all these cases, the analyte molecules link the nanoparticles which results in a plasmonic chiral optical response due to the prolate shape of the nanoparticles. This concept, however, relies on random defects of nominally highly symmetric nanoparticles for the generation of plasmonic chirality. The sensitivity can be largely increased if either particles of lower symmetry, such as rods, or intrinsically chiral assemblies are used. Ma and co-workers used the DNA triggered assembly of twisted chains of gold nanorods for DNA detection (*105*). The importance of the orientation of the nanorods has been pointed out by Han and collaborators, who showed that the side-by-side arrangement leads to stronger responses than an end-to-end configuration (*106*). Wen and collaborators detected copper ions in L-cysteine and gold intrinsic three-dimensional structure, that is, a pyramidal grouping of four nanoparticles. In the first study the presence of the target DNA molecule triggered a structural alteration and thus an increase in the chiral optical response. In the study of Oh and collaborators, a helical arrangement of gold nanoparticles was achieved (*107*). Li et al. used a combination of gold nanoparticles and luminescent particles that even allowed for a double detection, monitoring CD as well as the luminescence. It has to be mentioned that this indirect detection scheme is, in contrast to the previously discussed interaction of plasmonic and molecular resonances, not generally applicable. Only analytes that trigger the corresponding self-assembly process can be detected. However, if such a process exists, the resulting enhancement is beyond reach for any other method known thus far.

**Conclusions**

Chirality, in a very general sense, refers not only to a geometrical property but also to the optical response of a particular system. While the handedness of a system, from a purely geometrical point of view, constitutes a per se intriguing concept, a chiral optical response is a highly desirable property. While those two concepts are inextricably connected in molecular systems, chiral optical responses of plasmonic systems result from the handedness of the underlying plasmon modes, which not necessarily follow the structural symmetry of the plasmonic system.

Advances in micro- and nanofabrication techniques have enabled the realization of sophisticated chiral plasmonic systems. Direct laser writing, multi-layer lithography, self-assembly via scaffolds or DNA, and many more have been used to study the working principle of plasmonic chirality. In order to truly benefit from the strong light-matter interaction of plasmons collective mode formation has been identified as a key feature. Resonant interaction between adjacent nanoparticles enables a strong chiral optical response, much larger than in molecular systems and even large enough to create circular polarizers. Sophisticated materials and designs have furthermore lead to active chiral systems where the chiral optical response or even the handedness of the structure itself could be switched. The combination of plasmonic systems with chiral molecules shows promising prospects for enhanced enantiomer detection. With this rich set of fascinating results, chiral plasmonics has lived up to many of the original promises and hopes and will most definitely remain an interesting and fruitful field of research for the years to come.

**Associated Content**

**Materials and Methods**

None, review paper.

**Author Information**

Corresponding Author: Na Liu, na.liu@kip.uni-heidelberg.deCompeting Interests: The authors declare that they have no competing interests.

Author contributions: All authors contributed to the discussion, outline, and writing of the manuscript.

**Data and materials availability:** All data needed to evaluate the conclusions in the paper are present in the paper. Additional data related to this paper may be requested from the authors.

**Acknowledgments**

We gratefully acknowledge financial support by the Deutsche Forschungsgemeinschaft (SPP1391, FOR730, and GI 269/11-1), by the Bundesministerium für Bildung und Forschung (13N9048 and 13N10146), by the Baden-Württemberg Stiftung, by the Ministerium für Wissenschaft, Forschung und Kunst Baden-Württemberg (Az: 7533-7-11.6-8), Zeiss Foundation, Alexander-von-Humboldt Foundation, by the European Research Council (ERC Advanced Grant Complexplas), and by the German-Israeli Foundation. We also acknowledge financial support by the Sofja Kovalevskaja grant from the Alexander von Humboldt-Foundation, the Marie Curie CIG grant, and the European Research Council (ERC *Dynamic Nano*) grant.